Theoretical and Experimental Assessment of Thinned Germanium Substrates for III-V Multijunction Solar Cells

Iván Lombardero[1*], Mario Ochoa[1†], Naoya Miyashita[2], Yoshitaka Okada[2], Carlos Algora[1]

Instituto de Energía Solar – Universidad Politécnica de Madrid

ETSI de Telecomunicación, Avda. Complutense 30, 28040, Madrid, Spain

[1]Instituto de Energía Solar - Universidad Politécnica de Madrid, ETSI de Telecomunicación, Avda. Complutense 30, 28040 Madrid, Spain

[2]Research Center for Advanced Science and Technology (RCAST), The University of Tokyo, 4-6-1 Komaba, Meguro-ku, Tokyo 153-8904, Japan

*** Corresponding author.** ivan.lombardero@ies.upm.es (I. Lombardero)

† Present address: Laboratory for Thin Films and Photovoltaics, Empa-Swiss Federal Laboratories for Materials Science and Technology, Ueberlandstrasse 129, 8600 Duebendorf, Switzerland

***Abstract.*** Solar cells manufactured on top of Ge substrates suffer from inherent drawbacks that hinder or limit their potential. The most deleterious ones are heavy weight, high bulk recombination, lack of photon confinement and an increase of the heat absorption. The use of thinned Ge substrates is herein proposed as a possible solution to the aforementioned challenges. The potential of a thinned Ge subcell inside a standard GaInP/Ga(In)As/Ge triple-junction solar cell is assessed by simulations, pointing to an optimum thickness around 5-10 µm. This would reduce the weight by more than 90 %, whereas the available current for the Ge subcell would decrease only by 5 %. In addition, the heat absorption for wavelengths beyond 1600 nm would decrease by more than 85 %. The performance of such a device is highly influenced by the front and back surface recombination of the p-n junction. Simulations remark that good back surface passivation is mandatory to avoid losing power generation by thinning the substrate. In contrast, it has been found that front surface recombination lowers the power generation in a similar manner for thin and thick solar cells. Therefore, the benefits of thinning the substrate are not limited by the front surface recombination. Finally, Ge single-junction solar cells thinned down to 85 µm by wet etching processes are demonstrated. The feasibility of the thinning process is supported by the limited losses measured in the current generation (less than 6 %) and generated voltage (4 %) for the thinnest solar cell manufactured.

# 1   INTRODUCTION

Germanium solar cells (mostly used as multijunction's bottom subcell) are usually fabricated by metalorganic vapor phase epitaxy (MOVPE) on p-type Ge substrates. The emitter thickness is usually between 150 nm (for single-junctions) and around 200-250 nm (for triple-junctions, due to the extra thermal load),[1] made by the diffusion of P atoms into the substrate during the growth of the GaInP nucleation layer.[2,3] The rest of the Ge substrate (around 175 µm) works as the solar cell base. GaAs nucleation layers, which have also been reported,  are less frequent since they result in thicker emitters caused by the higher diffusivity of As in Ge.[4,5] Consequently, the substrate plays a key role as most of the absorption (and consequently the carrier collection) takes place in it. However, semiconductor substrates have a minimum thickness imposed by the manufacturing process (wafer cut out of the ingot and subsequent polishing) which implies some inherent drawbacks, hindering the solar cell performance, or at least, limiting its potential in favor of substrate-free solar cells. The main impediments caused by thick Ge substrates are:

- Heavier solar cells: Ge substrates accounts for 95 % of the weight in a triple-junction solar cell (detailed in Section 3.1). Moreover, the weight of a solar cell is a key feature for some markets where the key metric is not only $/W_P$ but $W/m^2$ or $W_P/kg$[6,7] (such as in space applications[8–10] or other niche markets).

- Higher operation temperature: highly doped Ge substrates (as the ones used for most solar cells) suffers from free-carrier absorption (FCA).[11–13] This type of absorption hinders the photogeneration for long wavelengths and extends the absorption beyond the bandgap (without generating electron-hole pairs) thus, heating up the solar cell.

- Lower open circuit voltage: Once the substrate is thick enough to absorb all the desired light, any extra thickness only contributes to increasing the bulk recombination.

- Lack of photon confinement: Conventional Ge substrates avoid the benefits of a longer optical path caused by a back reflector.[14,15] The two main reasons behind this phenomenon are the minority-carrier diffusion length, which is shorter than the substrate thickness, and the hindrance in the photogeneration caused by the FCA. Thinning the substrate would allow gathering the carriers generated by the light reflected in the back reflector.

In this work, we propose the use of thinned Ge substrates to solve, or at least mitigate, these detrimental effects. Although the development of Ge solar cells has been ongoing for years,[3,16,17] the use of thin substrates has not been thoroughly assessed yet. Even if some works claim to produce thinned Ge solar cells,[7] there is almost neither information about how to carry out the thinning process nor studies analyzing the performance of Ge solar cells with different substrate thicknesses. Furthermore, most of the efforts for the development of III-V multijunction solar cells have been focused on GaAs and GaInP subcells,[18] or in developing new materials[19–21] to increase the number of junctions. Conversely, thinning the Ge substrate would provide a way to improve the current technology (lighter devices, operating at lower temperatures and avoiding most of the bulk recombination) without involving major changes in the semiconductor structure nor the device design either. Moreover, this would permit the use of a robust, well-known structure such as the GaInP/Ga(In)As/Ge triple-junction solar cell for low mass applications, or even flexible ones if sufficiently thin devices are developed.

Substrate removal or epitaxial lift-off (ELO) techniques are currently employed on inverted metamorphic (IMM),[22–25] GaAs,[26] or GaInP/GaAs[27] solar cells to manufacture thin solar cells. The wafer spalling technique has also been reported as a promising manufacturing process for thin single and multijunction solar cells.[28,29] Other approaches focus on the epitaxial growth of thin Ge solar cells[30,31], exfoliate Ge substrates,[32] or the so-called Germanium-on-Nothing (GeON)[33,34], although neither of them has achieved yet the state-of-the-art of Ge solar cells. Despite the numerous possibilities, none of the aforementioned technologies has been able yet to become the workhorse for light-weight applications (an excellent review regarding the economic potential of ELO, spalling and GeON, considering the possibility of reusing the substrate, was published by J. Ward et al[35]).

Therefore, the main purpose of this work is to study all the benefits and challenges of thinning the Ge substrate and show a procedure to carry it out. First, we evaluate the performance of a thinned Ge solar cell inside a standard triple junction (3J) made of GaInP/Ga(In)As/Ge. Then, a wet etching process able to thin down Ge substrates is demonstrated and used to manufacture thinned Ge solar cells. Finally, a complete characterization of the solar cells has been carried out proving the feasibility of the thinning process.

## 2   METHODS

The device performance is assessed by 2D simulations carried out with Silvaco Atlas.[36,37] Neither shadowing nor resistive losses at the ohmic contacts have been taken into account. Optical calculations use the Transfer Matrix Method (TMM),[38] considering no antireflection coating (ARC) and a rear contact made of gold as a back reflector. The FCA is calculated using extinction coefficients reported in the literature.[13] Then, the photogeneration is corrected accordingly applying a wavelength-dependent factor as suggested elsewhere.[11,12] Changes in the operating temperature (due to a lower absorption as the substrate is thinned down) are not considered, as it will be highly dependent on the encapsulation and operating conditions of each device in particular.[39,40] Consequently, the temperature is set to 25 °C. The spectrum used to illuminate the solar cell in all simulations throughout this paper is the AM0 spectrum (136.6 mW/cm$^2$).

The Ge solar cell considered is a state-of-the-art-device with a $6 \cdot 10^{17}$ cm$^{-3}$ constant doping level for the base. The emitter doping profile has been assumed to be uniform at $10^{19}$ cm$^{-3}$, as a reasonable approximation.[1,41] Minority-carrier mobilities and lifetimes reported in the literature[42,43] are used as input for the simulations. The resultant minority-carrier diffusion length is around 115 μm in the base ($\tau_{e^-}$ = $8.2 \cdot 10^{-5}$ s, $\mu_{e^-}$ = 1019 cm$^2 \cdot$V$^{-1} \cdot$s$^{-1}$) and around 5 μm in the emitter ($\tau_{h^+}$ = $9.5 \cdot 10^{-8}$ s, $\mu_{h^+}$ = 116 cm$^2 \cdot$V$^{-1} \cdot$s$^{-1}$). These values are also supported by previous modelling of germanium solar cells.[1,3,44] A degradation of the window/emitter interface properties is usually reported as a consequence of the Ga diffusion from the upper layers towards the emitter.[1,41] Accordingly, an interface trap density of $5 \cdot 10^{11}$ cm$^{-3}$ with a capture cross section of $5 \cdot 10^{-15}$ cm$^2$ was set at the window/emitter interface. This results in a surface recombination of $2 \cdot 10^5$ cm/s, similar to what has been reported in the literature.[3]

Dirichlet boundary conditions are used at the contacts for which the quasi-Fermi level potentials equal the voltage applied. An implication of this assumption is that all carriers are effectively recombined at the contacts, i.e. infinite surface recombination velocity. Therefore, to be able to modify the back surface recombination (BSR) velocity, and only for simulation purposes, the infinite recombination at the rear contact has to be avoided. An ideal electron collector[45] has been implemented by a heterojunction at the back surface made of a 20 nm-thick 1 eV bandgap material (with material properties similar to those of Ge except for the bandgap energy and affinity). The electron affinity of this ideal "ad-hoc" material has been set to produce no band offset at the valence band, creating a 375 meV barrier at the conduction band to effectively block the minority carrier transport to the rear contact. Then, the desired recombination velocity for electrons and holes can be set at the interface between the electron collector and the base. Infinite BSR velocity is assumed to be simulated for recombination velocities higher than $10^9$ cm/s, as it reproduces a similar behavior to the one obtained in the absence of our ideal electron collector.

Germanium single-junction solar cell structures were grown in a commercial Aixtron 200/4 MOVPE reactor on p-type Ge(100) wafers gallium doped $6 \cdot 10^{17}$ cm$^{-3}$ (measured by ECV), a miscut of $6^0$ (measured by HRXRD) towards the nearest [111] and a thickness of 175 µm. An 80 nm thick GaInP nucleation layer on Ge was employed, followed by a 680 nm thick Ga(In)As cap layer to facilitate the formation of ohmic contacts. Details on the growth can be found elsewhere.[46] The substrate manufacturer provides the substrate thickness with an uncertainty of ±10 µm (i.e. 165-185 µm). Unless otherwise noted samples are assumed to be 175 µm thick for the sake of simplicity. Ge substrates were thinned using a wet etching process. A more detailed explanation can be found in Section 4.1.

Solar cells with an active area of 0.09 cm$^2$ were manufactured by standard photolithography techniques and thermal evaporation for the formation of the metal contacts (AuGe/Ni/Au was employed for the front contact metallization and just Au for the rear one). No ARC was deposited on the devices.

External Quantum Efficiency (EQE) measurements were taken using a custom-built system based on a 1000-W Xe-lamp and a triple-grating *Horiba Jobin Yvon* monochromator (TRIAX180), equipped with a filter wheel and monitor sensors to account for the light source instabilities. The corresponding signal has been detected by an *eLockIn 203 dual-input 4-phase DSP* lock-in amplifier detector in conjunction with an *SR570* transconductance preamplifier for each channel.

J-V curves under illumination were measured with no spectral adjustment using the 4-point probe technique on a temperature-controlled chuck using a 4-quadrant source-monitor unit. Dark J-V measurements were taken under similar conditions while avoiding any source of light by means of an enclosure that totally covers the measurement setup. Transmittance (*T*) and reflectance (*R*) of the bare substrates were characterized by a PerkinElmer Lambda 1050 spectrophotometer.

## 3   MODELING OF THINNED GE SOLAR CELLS

The use of highly doped substrates (around $10^{17}$-$10^{18}$ cm$^{-3}$) is wide-spread for GaInP/Ga(In)As/Ge triple-junction solar cells.[1,3] In this structure, the bottom subcell generates an excess of current in comparison with the upper subcells. Consequently, highly doped substrates are commonly used as they increase the $V_{OC}$ and the *FF* without limiting the current generation. Although the use of low doped substrates has been reported for some applications in particular, such as for thermophotovoltaic (TPV)[17] or upright metamorphic (UMM)[47] solar cells (due to their lower FCA and longer diffusion length[42,43]), their use is very limited for photovoltaic applications. Regarding the substrate thickness, it is usually around 165-185 µm to ensure a high throughput during the substrate manufacturing process[48]. Accordingly, we focused our study on typical Ge solar cells grown on 175 µm thick substrates p-doped 6·$10^{17}$ cm$^{-3}$.

### 3.1  DECREASE OF THE SOLAR CELL WEIGHT

Looking at the main component materials in a multijunction solar cell (GaInP, GaAs and Ge) and at their respective densities (4.47, 5.32 and 5.32 g·cm$^{-3}$) it can be seen that Ge accounts for more than 95 % of the total weight (see FIGURE 2). 2 µm of GaInP, 5 µm of GaAs and 175 µm of Ge are assumed for this calculation (GaInP and GaAs solar cell thicknesses are overestimated to take into account other layers apart from the subcells themselves, such as the tunnel junctions or the buffer layers).

To evaluate how much the substrate can be thinned down without limiting the current generation, the available current for the bottom subcell is depicted as well in FIGURE 2. This calculation considers that only wavelengths between 885 nm (GaAs absorption) and 1600 nm (due to the FCA) are useful for the Ge subcell, assuming 100 % collection efficiency. On the other hand, the current available for the GaInP and GaAs subcells is 19.8, 16.4 and 14.5 mA/cm$^2$ under AM0, AM1.5G and AM1.5d spectra respectively (between 300 and 885 nm assuming 100 % collection). Although this is a very simplified calculation (reflection and recombination losses have not been taken into account), it gives a rough idea about the minimum thickness needed for the Ge substrate in a 3J solar cell.

FIGURE 2 shows that the available current for the bottom subcell starts to decrease once the weight-loss starts to saturate. The thickness range between 5-10 μm is of particular interest as both, the available current and the weight-loss are above 90 %. Moreover, the available current for the upper subcells is still well below the available one for the bottom subcell, even if it were thinned down to 5 μm, ensuring a good performance of the thinned Ge in a 3J.

## 3.2  HEAT GENERATION DIMINISHING

FCA increases the intra-band absorption beyond 1600 nm, extending it to the whole solar spectrum. These wavelengths (λ > 1600 nm) account for 10 % of the total irradiance impinging the solar cell under AM0 spectrum and about 8 % under AM1.5G and AM1.5d spectra. Moreover, they mean 13.9 % (13.3 % under concentration assuming 100 % transmittance of the optic elements) of the total energy heating up the solar cell if we consider that the rest of the irradiance is transformed into electricity with an efficiency higher than 28 % (40 % under concentration) by a standard 3J. The heat absorption (considered as the wavelengths absorbed beyond 1600 nm) as a function of the substrate thickness has been assessed simulating the internal absorptance to avoid the influence of any optical effects:

$$A_{Internal}(\lambda) = \frac{A(\lambda)}{\left(1 - R(\lambda)\right)} \qquad (1)$$

where $A$ stands for the absorptance and $R$ for the reflectance. Optical calculations were performed using the TMM with reported refraction indexes for p-type doped Ge.[13] FIGURE 3 shows that only wavelengths absorbed by the FCA are decreased while the substrate is thinned down up to 50 μm. Then, wavelengths shorter than 1600 nm (i.e. wavelengths that contribute to the photogeneration) start to decrease as well. Nevertheless, the aforementioned 5-10 μm thickness range (see Section 3.1) would decrease the FCA from 60 % to less than 8 %. Accordingly, the heat absorption for long wavelengths can be decreased by more than 85 % using thinned Ge subcells without limiting the current generation in a 3J solar cell. The impact of the lower heat generation on the operating temperature of the solar cell must be calculated considering the operating conditions (spectrum irradiance, ambient temperature, etc.) and the device packaging (solar cell size, thermal dissipation block, etc.) in a given solar panel.[39,40]

## 3.3  PERFORMANCE AT 1 SUN

Ge subcell I-V curves under AM0 spectrum (136.6 mW/cm$^2$) filtered by a GaAs layer (in order to emulate the light absorption of top and middle subcells in a multijunction) have been simulated to assess the influence of the substrate thickness. An n/p Ge single-junction has been simulated instead of the 3J (GaInP/Ga(In)As/Ge), to simplify the calculations (similar to the manufactured one, see FIGURE 1, except for a thicker GaAs cap layer of 5 μm thickness instead of 680 nm, to act as a GaAs filter). The upper subcells have been omitted because their performance can only be influenced by a thinner substrate if the operation temperature changes (as long as the Ge subcell does not limit the current generation) which is not considered in our model (see Section 2). One key parameter to evaluate together with the substrate thickness is the recombination at the rear contact. Its influence is usually neglected because, for standard Ge solar cells, the diffusion length of electrons[42,43] does not allow a high minority carrier population at the back surface. However, its deleterious effects will become increasingly noticeable as the substrate is thinned down.

FIGURE 4 depicts the dependence of $J_{SC}$, $V_{OC}$, $FF$, and $P_{Max}$ with the substrate thickness under four different boundary conditions: no surface recombination, only standard window/emitter front surface recombination (FSR), only infinite BSR, and combination of standard FSR and infinite BSR. In the absence of any surface recombination, thinning the substrate increases the $V_{OC}$ and the $FF$ while decreasing the $J_{SC}$, as expected. The highest $P_{Max}$ is achieved around 10 μm for this ideal scenario. Looking at the influence of the surface recombination, thick substrates are clearly dominated by the FSR while being insensitive to the back one. The $J_{SC}$ is reduced by an offset in the presence of FSR with respect to the ideal case while the improvement of the $V_{OC}$ and the $FF$ is limited. Conversely, thin substrates are dominated by the BSR as it affects $J_{SC}$, $V_{OC}$, and $FF$ while the presence of FSR only translates into an offset in these parameters (in the presence of BSR), which is almost negligible for $V_{OC}$ and $FF$ in very thin substrates. This different behavior is explained as the emitter thickness is constant and it does not change as the substrate is thinned down. On the other hand, the thinner the substrate the closer the back surface is to the *pn* junction, increasing its influence on the overall performance. This points out that it is more important to avoid the back surface recombination than the front one in order to boost the development of thinned Ge solar cells.

Regarding the performance in a 3J, it must be considered that state-of-the-art devices generate more than 17 mA/cm$^2$ under AM0 spectrum.[7] This value could be achieved under all boundary conditions in the aforementioned thickness range of interest (5-10 µm), except for the one with high recombination on both surfaces. Nevertheless, our simulations do not take into account the use of an ARC which usually increases the current at least by 15 %, which would allow generating more than 18.5 mA/cm$^2$ even in the worst case of high recombination in both surfaces. Looking at the performance, it stands out that it does not increase except for the scenario with no surface recombination. Nevertheless, this does not necessarily mean that thinning the substrate is of no interest. Even in the case of a lower performance, it could be a profitable choice in exchange for a lighter device operating at a lower temperature for certain applications. Moreover, the cooler operation temperatures could counterbalance the performance decrease caused by the thinner substrate. It should also be considered that the current limitation in a 3J implies that the maximum power depicted for the Ge subcell does not correspond to the actual one in a 3J. The Ge subcell will be operating at the minimum current generated in the 3J and therefore, the power generation in the Ge subcell will be insensitive to its current loss as long as it is not the limiting subcell. This is depicted in FIGURE 5, where the power generation is recalculated assuming that the current generation is limited to 17 mA/cm$^2$ causing no *FF* nor $V_{\text{PMAX}}$ variation (i.e. no change in the I-V operation point of the Ge subcell). Under such an assumption, the FSR lowers its influence even more, allowing for 3 µm thick Ge subcells with no power generation degradation in the absence of BSR. Furthermore, in this scenario the power generation will be maximized at 15 µm. The presence of high BSR is still highly deleterious, although its effects are slightly mitigated until the Ge subcell becomes the limiting subcell. In contrast, in the absence of surface recombination the power generation improvement has been limited, and the optimum thickness is thinner (2.5 µm) than in the previous case. These results reinforced the previous conclusion pointing to the BSR as the key parameter to allow developing thinned Ge subcells.

Consequently, we proceed to evaluate the back surface recombination in the Ge subcell performance. FIGURE 6 depicts two contour plots (with and without considering a 2·10$^5$ cm/s FSR) of the maximum power point as a function of the thickness and the BSR velocity (no current limitation inside a 3J has been considered in this case). It shows that the BSR velocity should be limited below 10$^2$ cm/s to completely avoid its influence on the solar cell performance for both scenarios, although a recombination velocity of $10^3$ cm/s only degrades the performance around 0.05 mW/cm$^2$ in the instance of high FSR (in the absence of FSR the power degradation due to the BSR is highly dependent on the solar cell thickness). Therefore, efforts improving the BSR should have this value ($10^3$ cm/s) as a minimum target disregarding the FSR. Silicon-based passivation layers and aluminum-diffused back surface field (BSF) layers have been proposed as means to decrease the BSR velocity in p-type Ge for TPV and space solar cells,[17,49,50] achieving recombination velocities as low as 17 cm/s in bare substrates (i.e. without metal contacts).[51] Therefore, it seems realistic to achieve such a low BSR velocity, although a proof of concept of a good enough back surface passivation together with a metal contact is still to be demonstrated.

## 4  EXPERIMENTAL

Once the advantages of a thinner Ge substrate have been assessed we describe our thinning process and evaluate the performance of the manufactured solar cells.

## 4.1 GERMANIUM THINNING PROCESS

Chemical etching has been selected as the method to thin down the Ge substrate as it is scalable, easy to apply and uses standard materials or tools in the manufacturing process of solar cells. Moreover, it has been extensively studied in Ge substrates, being plenty of options available to carry it out.[52–56] Although most of the previous studies focused on etching depths much lower than our interests (10 µm or less), $H_3PO_4$:$H_2O_2$:$H_2O$ proved to be a good candidate for long etching processes. A good trade-off between the etch rate and the surface homogeneity was achieved for a 1:6:3 ratio.[56] Moreover, certain studies used the same chemical agent to carry out the Ge hydrolysis from certain compounds (such as magnesium germanide or magnesium silicide) to recycle Ge. This opens the door to recycle the etched Ge (although further study is required to properly evaluate this possibility) which would aid ensuring the supply of this strategic material in the long term.[57–59] Alternative processes to recycle Ge from the manufacturing process of multijunction solar cells on Ge substrates are ultra-purification, reverse osmosis and evaporation.

The depth achieved by the etching process was assessed as follows. First, a photolithography process was performed to avoid etching some areas of the sample. Second, the sample was put in the solution with the side to be etched facing up. Once the etching process was completed, the sample was taken out and the photoresist was removed. Finally, the etched profile was measured by means of a profilometer. All the etching processes were carried out at ambient temperature (23-27 °C) with enough solution in order to avoid its saturation. Etching processes up to 9 hours were carried out with an almost constant etching rate around 10 μm/hour (FIGURE 7), similar to what has been reported in the literature.[56] Thinner samples have not been manufactured yet due to the increasing fragility as the substrate thickness decreases. Currently, a new manufacture process is under development to allow thinning substrates down to even thinner thicknesses.

## 4.2  SOLAR CELL CHARACTERIZATION

Ge single-junction solar cells (FIGURE 1) were thinned down for 3, 6 and 9 hours (obtaining 145, 115 and 85 μm thick semiconductor structures) keeping one extra sample unetched (not thinned at all, i.e. 175 μm) for comparison. Bare substrates (i.e. without any semiconductor structure grown on top of it nor any metal deposited either) were thinned down for similar times to allow an easier optical characterization of the substrate absorption. First, transmittance ($T$) and reflectance ($R$) of the bare substrates were characterized. Then, the absorptance ($A$) was calculated as $A = 1 - R - T$. Finally, the internal absorptance was calculated using Equation (1), obtaining the results depicted in FIGURE 8. It shows that the internal absorptance remains constant at 100 % for wavelengths shorter than 1600 nm, even for the thinnest sample, while decreasing for longer wavelengths. The difference in the internal absorptance between the 175 and the 85 μm thick samples is an almost constant 18 % for wavelengths longer than 1600 nm. This is in agreement with simulations that pointed to an absolute difference around 20.5 % (see FIGURE 3, also plotted in FIGURE 8 for comparison purposes). Moreover, the internal absorptance is well fitted for the 85 nm thick substrate. On the other hand, the 115 and 145 μm internal absorptance is slightly underestimated (differences < 3 %) while it is overestimated for the 175 μm thick sample (still < 3 %). Nonetheless, the 175 μm internal absorptance is anomalously low between 1700 and 2000 nm, as it is similar to the one measured at 145 μm, being as expected (higher internal absorptance) for longer wavelengths. These variations fall within the thickness uncertainty of the substrates, and the measurements follow the overall trend. Consequently, the results support the presence of FCA in Ge substrates and the thinning process as a valid solution to mitigate it.

Then, the J-V curve under illumination (AM0 – 136.6 mW/cm$^2$) has been measured for all samples (FIGURE 9). A small degradation can be observed for both, $J_{SC}$ and $V_{OC}$, consequence of an increasing influence of the back surface recombination at the back metal contact. Nevertheless, it is demonstrated that a germanium single-junction solar cell without any back surface passivation can be thinned down to 85 µm losing only 6 % $J_{SC}$ and 4 % $V_{OC}$. The *FF* degradation is lower than the influence of the number of fingers of the front metal grid (lower than a relative 5 %) and therefore, a trend is hardly seen. As a result, the power generation decreases by 6 %, being only 4 % in a 3J (the Ge subcell still generates a notable excess of current in comparison with the 3J and therefore, only the $V_{OC}$ loss will influence its performance in a 3J). These results show that the simulations (see FIGURE 3) have slightly underestimated the degradation measured in the solar cells. Although the trend is similar, currents 4.5 % and voltages 3.5 % lower than expected have been measured. This could be explained by a red-rich spectrum in the solar simulator or by a lower reflectivity than expected at the rear surface, although further work is needed to explain these differences. To further confirm these results dark J-V measurements have been carried out (FIGURE 10). They show an increase in the recombination currents, being dominated by a *kT* voltage-dependent recombination (i.e. n=1) for all samples. Looking at an equivalent current to the $J_{SC}$ in the dark J-V and comparing the voltages to the $V_{OC}$ (see insets in FIGURE 9 and FIGURE 10) a good agreement is achieved with differences lower than 50 mV, which further confirms the previous results. FIGURE 11 shows the EQE measurements for each sample, following a similar trend than the J-V curves, decreasing with the substrate thickness. The degradation is mostly observed at long wavelengths in agreement with simulation results (not shown for the sake of simplicity). The current loss calculated with the EQE under AM0 spectrum is around 4% between the 175 and the 85 µm thick samples. This is in reasonable agreement (considering the sample-to-sample variation) with the illumination I-V curve measurements, confirming the previous results.

## 5   CONCLUSIONS

Through this article the deleterious effects caused by the substrate in Ge solar cells have been pointed out, namely, heavier solar cells operating at a higher temperature, generating lower voltages and unable to benefit from a back reflector. Thinning the substrate down to 5-10 µm would reduce the weight by more than 90 % while limiting the available current loss below 10 % and reducing the heat absorption in unusable wavelengths by more than 85 %. The back surface recombination has been detected as a key parameter for thinned Ge substrates. Recombination velocities below $10^3$ cm/s should be achieved to avoid its detrimental effect. On the other hand, even if the Ge subcell suffers from high front surface recombination, thinning the substrate can still avoid unnecessary weight and heat absorption while generating a similar amount of power.

$H_3PO_4$:$H_2O_2$:$H_2O$ at a 1:6:3 ratio has been studied to thin down the substrate of a solar cell, obtaining constant etching rates about 10 µm/hour for etchings up to 90 µm (9 hours). Optical measurements on thinned substrates proved the thinning as a valid solution to decrease the FCA in Ge solar cells. Solar cells thinned down to 85 µm have demonstrated the feasibility of the manufacturing process, losing only 6 % of the current generation and 4 % of the generated voltage without any passivation layer nor BSF at the rear contact. This degrades the power generation of a Ge single-junction solar cell by 6 %, which in a 3J would only be 4 % as the Ge subcell will still generate an excess of current in comparison with the 3J.

## Acknowledgments

The authors would like to acknowledge the Silicon Device's group at CIEMAT for the optical characterization of the Ge substrates. We are also pleased to thank I. García and M. Hinojosa for dedicated work during the epitaxial growth of the Ge single-junctions by MOVPE and L. Cifuentes for fabricating the solar cells. This work has been supported by the Fundación Iberdrola España Research Grants, Spanish MINECO through the project TEC2017-83447-P and by the Comunidad de Madrid through the project MADRID-PV2 (S2018/EMT-4308). I. Lombardero gratefully acknowledges the financial support from the Spanish Ministerio de Educación, Cultura y Deporte through the Formación del Profesorado Universitario grant with reference FPU14/05272.

# LIST OF TABLES

# LIST OF FIGURES

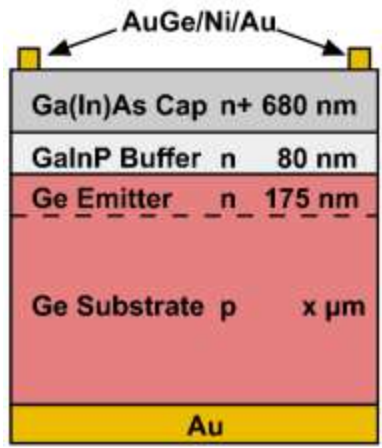

**FIGURE 1** Schematic layer structure of the solar cells analyzed in this work.

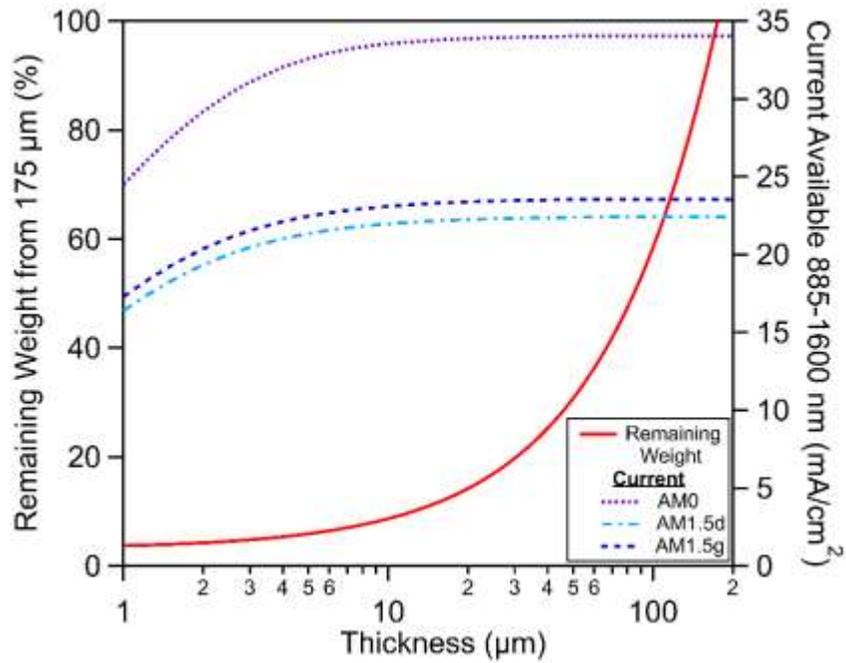

**FIGURE 2** Weight reduction of a 3J solar cell as a function of the Ge substrate thickness together with the available current for the Ge subcell between 885 nm (GaAs absorption) and 1600 nm (FCA) under AM0, AM1.5G, and AM1.5d spectra.

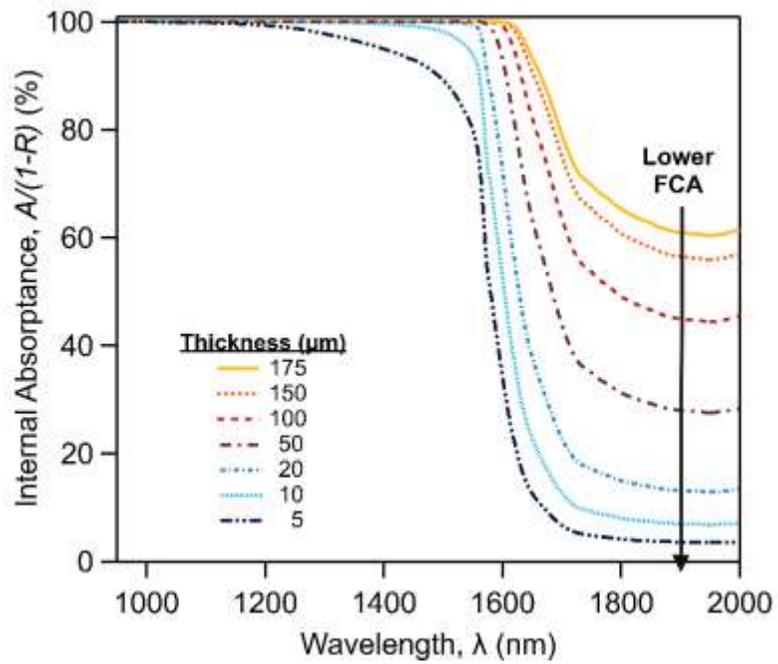

**FIGURE 3** Simulated internal absorptance (see Equation 2) as a function of the substrate thickness

of a Ge subcell inside a 3J solar cell. The substrate is assumed to be p-type doped $6 \cdot 10^{17}$ cm$^{-3}$.

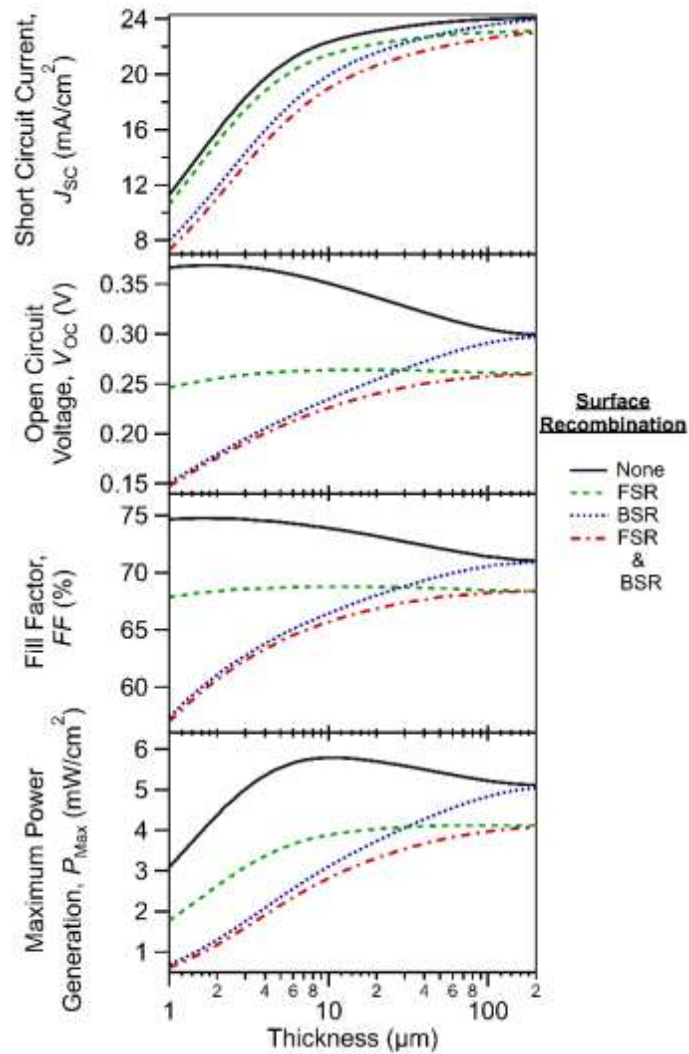

**FIGURE 4** $J_{SC}$, $V_{OC}$, $FF$ and $P_{Max}$ of a Ge single-junction solar cell filtered by a GaAs layer under AM0 spectrum (136.6 mW/cm$^2$) as a function of the substrate thickness. Four scenarios are plotted regarding the Ge solar cell surface recombination: *None:* no surface recombination, *FSR:* only standard window/emitter surface recombination, *BSR*: only infinite back surface recombination and *FSR & BSR:* combination of standard window/emitter and infinite back surface recombination. Infinite recombination is assumed to be $10^9$ cm/s.

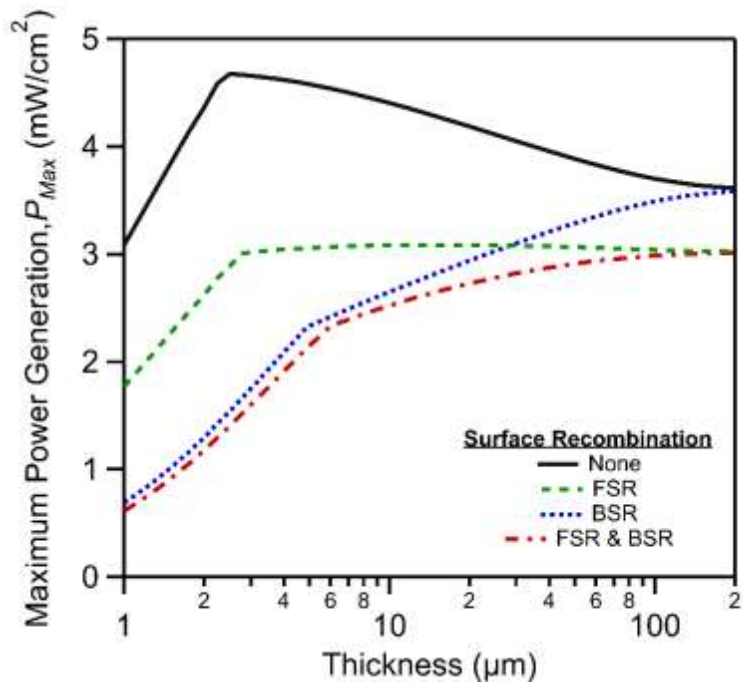

**FIGURE 5** $P_{Max}$ of a Ge subcell inside a 3J solar cell under AM0 spectrum (136.6 mW/cm²) as a function of the substrate thickness (current generation is considered to be limited to 17 mA/cm² by the top/middle subcells). Four scenarios are plotted regarding the Ge solar cell surface recombination: *None*) no surface recombination, *FSR*) only standard window emitter surface

recombination, *BSR*) only infinite back surface recombination and *FSR & BSR*) combination of standard window/emitter and infinite back surface recombination. Infinite recombination is assumed to be $10^9$ cm/s.

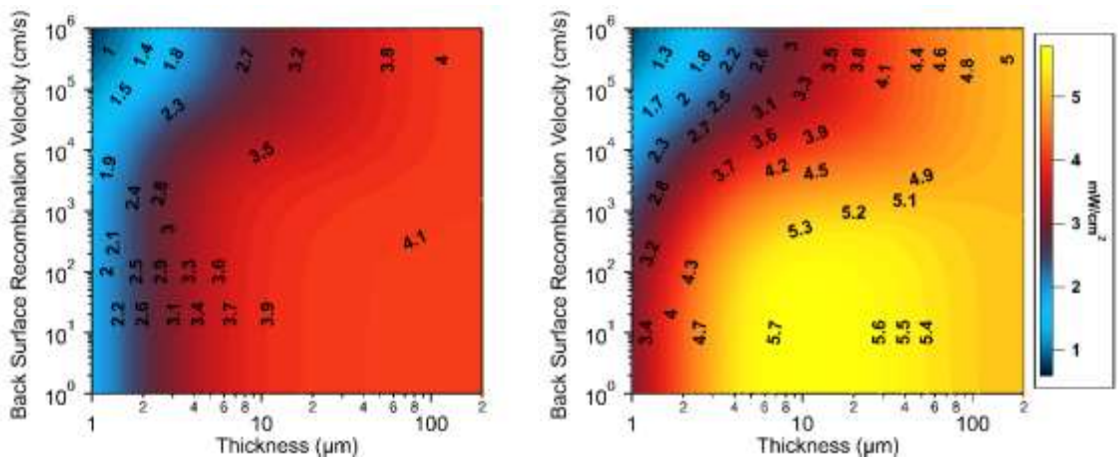

**FIGURE 6** $P_{Max}$ (mW/cm$^2$) generated by a Ge single-junction solar cell under AM0 spectrum (136.6 mW/cm$^2$) filtered by a GaAs layer, as a function of the substrate thickness and the back surface recombination velocity with (left) and without (right) considering a 2·10$^5$ cm/s FSR. The maximum BSR depicted is $10^6$ cm/s for clarity as the difference with $10^9$ cm/s is negligible. The cross-section of each contour plot at the minimum and maximum BSR velocities are indeed the four scenarios for the maximum power generation depicted in FIGURE 4.

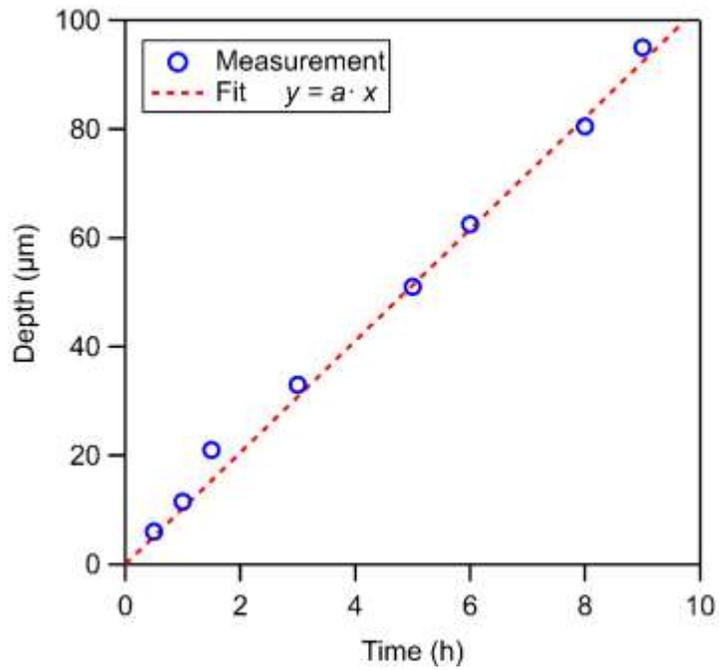

**FIGURE 7** Etched depth as a function of time for Ge substrates in $H_3PO_4$:$H_2O_2$:$H_2O$ at a 1:6:3 ratio.

A linear fit (y = a·x) is shown, being $a$ equal to 10.26 μm/h and $R^2$ equal to 0.995.

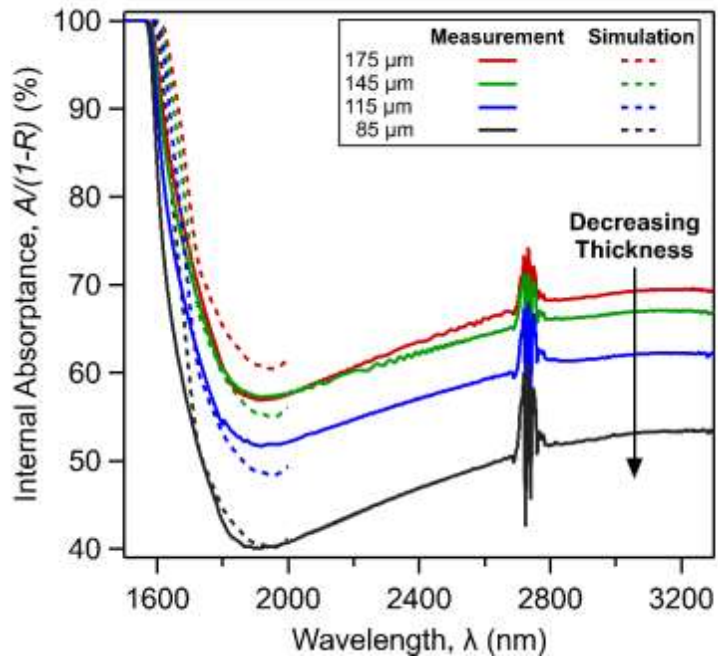

**FIGURE 8** Measured (solid) and simulated (dashed) internal absorptance (see Equation 2) of bare Ge substrates thinned for 3, 6 and 9 hours (145, 115 and 85 μm thick respectively). A standard substrate (175 μm) is also shown for comparison. The noise around 2750 nm is caused by water vapor absorption during the measurements (slight differences in the relative humidity in the laboratory).

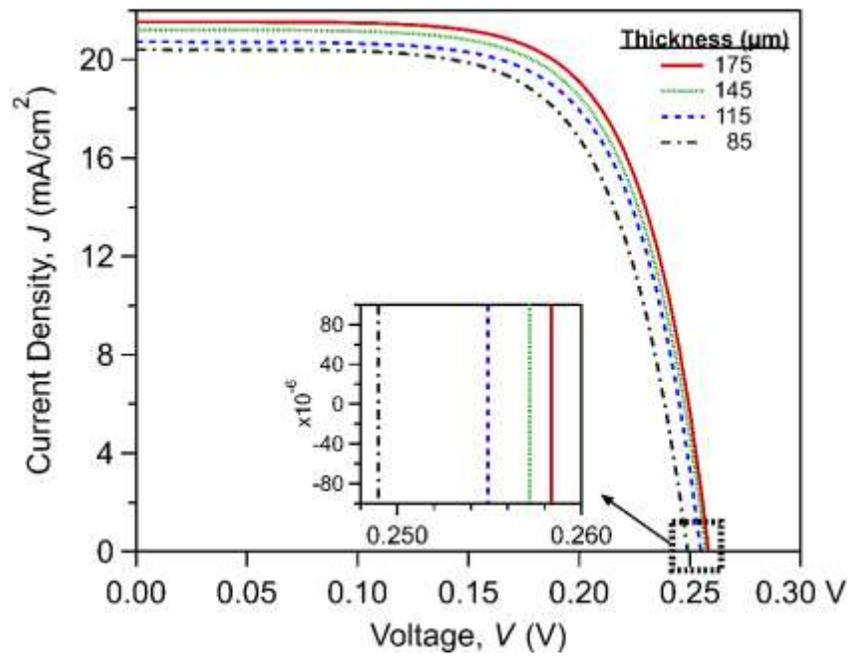

**FIGURE 9** J-V curves under illumination conditions (AM0 – 136.6 mW/cm$^2$) of Ge single-junction solar cells thinned for 3, 6 and 9 hours (145, 115 and 85 μm thick respectively). A standard single-junction (175 μm) is also shown for comparison.

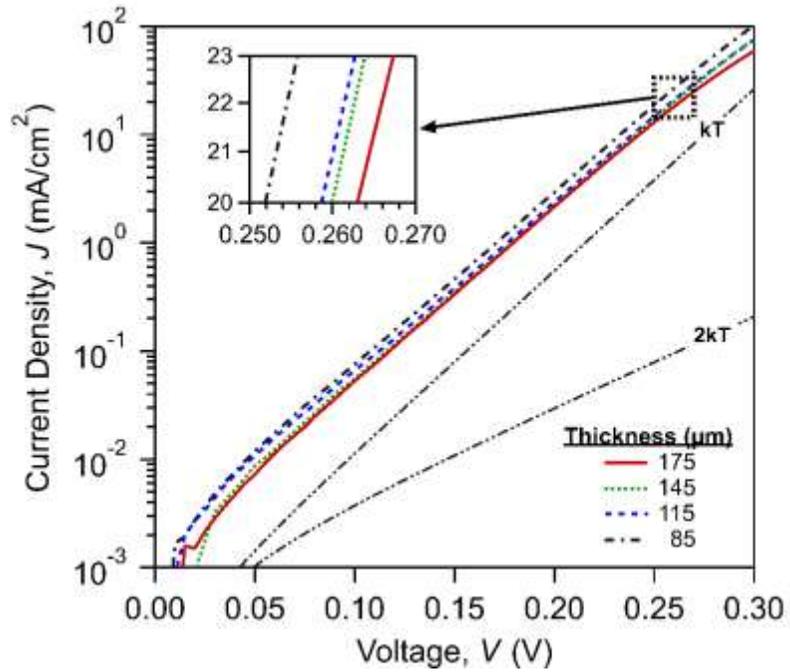

**FIGURE 10** Dark J-V curves of Ge single-junction solar cells thinned for 3, 6 and 9 hours (145, 115 and 85 μm thick respectively). A standard single-junction (175 μm) is also shown for comparison.

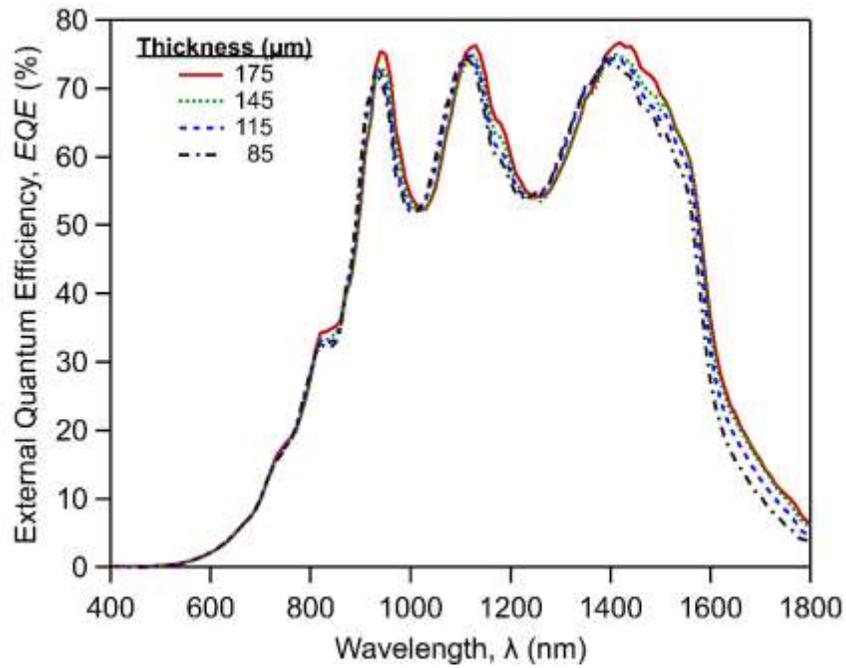

**FIGURE 11** EQE measurements of Ge single-junction solar cells thinned for 3, 6 and 9 hours (145, 115 and 85 μm thick respectively). A standard single-junction (175 μm) is also shown for comparison.